\documentclass[12pt,preprint]{article}
\usepackage{graphicx}

\newcommand\pubnumber{DPF2015-290}
\newcommand\pubdate{\today}

\def\napoli{Department of Physics\\
The Ohio State University, Columbus, OH 43210, USA}
\def\support{\footnote{Work supported in part by the Department of Energy under grant \uppercase{DE-SC}0011726.}}

\def\Title#1{\begin{center} {\Large #1 } \end{center}}
\def\Author#1{\begin{center}{ \sc #1} \end{center}}
\def\Address#1{\begin{center}{ \it #1} \end{center}}

\newcommand\pubblock{\rightline{\begin{tabular}{l} \pubnumber\\
         \pubdate  \end{tabular}}}
\newenvironment{Abstract}{\begin{quotation}  }{\end{quotation}}
\newenvironment{Presented}{\begin{quotation} \begin{center} 
             PRESENTED AT\end{center}\bigskip 
      \begin{center}\begin{large}}{\end{large}\end{center} \end{quotation}}

\def\beq{\begin{equation}}
\def\eeq#1{\label{#1}\end{equation}}
\def\eeqn{\end{equation}}

\def\beqa{\begin{eqnarray}}
\def\eeqa#1{\label{#1}\end{eqnarray}}
\def\eeqan{\end{eqnarray}}

\let\bar=\overbar

\def\Dslash{\not{\hbox{\kern-4pt $D$}}}
\def\dslash{\not{\hbox{\kern-2pt $\del$}}}

\def\msb{{\bar{\ssstyle M \kern -1pt S}}}

\begin{document}
\begin{titlepage}
\pubblock

\vfill
\Title{Inclusive Higgs Production at Large Transverse Momentum}
\vfill
\Author{ Hong Zhang, Eric Braaten\support}
\Address{\napoli}
\vfill
\begin{Abstract}
We present a factorization formula for the inclusive production of the Higgs boson 
at large transverse momentum $P_T$ that includes all terms with the leading power of $1/P_T^2$.
The cross section is factorized into convolutions of parton distributions, 
infrared-safe hard-scattering cross sections for producing a parton, 
and fragmentation functions that give the distribution 
of the longitudinal momentum fraction of the Higgs relative to the fragmenting  parton.
The infrared-safe cross sections and the fragmentation functions
are perturbatively calculable. 
The factorization formula enables the resummation of 
 large logarithms of $P_T/M_H$ due to final-state radiation 
by integrating evolution equations for the fragmentation functions.
By comparing the cross section for the process $q\bar{q}\to H t\bar t$ from the leading-power factorization formula 
at leading order in the coupling constants with the complete leading-order cross section, 
we infer that the error in the factorization formula decreases to
less than 5\% for $P_T>600$ GeV at a future $100$ TeV collider.
\end{Abstract}
\vfill
\begin{Presented}
DPF 2015\\
The Meeting of the American Physical Society\\
Division of Particles and Fields\\
Ann Arbor, Michigan, August 4--8, 2015\\
\end{Presented}
\vfill
\end{titlepage}
\def\thefootnote{\fnsymbol{footnote}}
\setcounter{footnote}{0}

\section{Introduction}

It is widely believed that the Higgs boson may reveal clues of the physics beyond the Standard Model. 
Among the basic properties of the Higgs are its production rates in high energy collisions. 
In hadron collisions, the Higgs is produced primarily with transverse momentum $P_T$ smaller than its mass $M_H$.
However its production rate at much larger $P_T$ is important, because it may be more sensitive to 
physics beyond the Standard Model, such as the decay of a much heavier particle into the Higgs.

The difficulty in calculating the Higgs $P_T$ distribution is a result of its multi-scale nature.
There are at least four scales in the calculation, including the partonic center-of-momentum energy $\sqrt{\hat s}$, $P_T$, $M_H$ and $M_i$, where $M_i$ is the mass of the particle from which the Higgs is emitted.
In regions where some of these scales are well separated, effective field theory may be used to separate these scales and simplify the calculation.
The most successful effective field theory in Higgs physics is the HEFT, in which virtual top-quark loops
are replaced by local interactions of the Higgs with  the vector bosons.
It is reliable in the region where $M_t$ is much larger than other scales in the problem.
HEFT is very successful in explaining the low $P_T$ distribution and the total cross section, which is dominated by the low  $P_T$ region.

However, it has been shown that HEFT is not effective in the large $P_T$ region.
Comparison with complete NLO calculations indicates that the difference between the exact and HEFT results  
exceeds 5\% at $P_T=150$ GeV and increases with $P_T$ \cite{Bagnaschi:2011tu}. 
The effects of the dimension-7 operators in HEFT on the Higgs $P_T$ distribution have recently been considered \cite{Harlander:2013oja, Dawson:2014ora}.
The expansion in the higher-dimension operators of HEFT breaks down for $P_T$ above about 150~GeV.

In this work \cite{Braaten:2015ppa}, we present the leading-power (LP) factorization formula for inclusive production of the Higgs at $P_T$ much larger than its mass $M_H$.
By separating the kinematic scales from the mass scales, the calculation of the Higgs $P_T$ distribution at large $P_T$ is greatly simplified.
The theoretical error can be decreased by summing logarithms of $P_T/M_H$ to all orders by integrating the evolution equations.

%%%%%%%%%%%%%%%%%%%%%%%%%%%
\section{Leading-power factorization formula}

The differential cross section for the inclusive production of a Higgs boson
in the collision of hadrons $A$ and $B$ can be written in a factorized form \cite{Collins:1989gx}:
%=========================
\begin{eqnarray}
\label{eq:Hfactorization}
{d}\sigma_{AB\rightarrow H+X}(P_A,P_B,P)
&=& 
\sum_{a,b}\int_0^1\! {d} x_a\, f_{a/A}(x_a,\mu) \int_0^1\! {d} x_b\, f_{b/B}(x_b,\mu)
\nonumber\\
&& \hspace{1cm}
\times
{d} \hat{\sigma}_{ab\to H+X}(p_a=x_aP_A,p_b=x_bP_B,P;\mu),
\end{eqnarray}
%=========================
where $P$ is the momentum of the Higgs, 
$P_A$ and $P_B$ are the momenta of the colliding hadrons, 
and $p_a$ and $p_b$ are the momenta of the colliding partons.
The sums in Eq.~(\ref{eq:Hfactorization})
are over the types of QCD partons, which consist of the gluon 
and the quarks and antiquarks that are lighter than the top quark.
The integrals in Eq.~(\ref{eq:Hfactorization}) are over the longitudinal momentum fractions 
of the colliding QCD partons.
The separation of the hard momentum scales of order $M_H$ and larger from the 
nonperturbative QCD momentum scale $\Lambda_{\mathrm{QCD}}$ 
involves the introduction of an intermediate but otherwise  arbitrary
factorization scale $\mu$.
The parton distribution functions $f_{a/A}$ and $f_{b/B}$
are nonperturbative, but their evolution with $\mu$ is perturbative.
Their evolution equations can be used to sum logarithms of
$M_H^2/\Lambda_{\mathrm{QCD}}^2$ to all orders in perturbation theory.
The integration range of $x_a$ is from 0 to 1,
up to kinematic constraints from  ${{d}}\hat{\sigma}_{ab\to H+X}$.

If the transverse momentum $P_T$ of the Higgs is much larger than its mass $M_H$, 
it is reasonable to expand the hard-scattering differential cross section 
$d\hat{\sigma}$ in Eq.~(\ref{eq:Hfactorization}) in powers of $1/P_T^2$.
The factorization formula for the leading power can be inferred 
from the perturbative QCD factorization formula for inclusive hadron production 
at large transverse momentum \cite{Collins:1989gx}.
The leading-power (LP) factorization formula for the 
hard-scattering differential cross section 
for the inclusive production of a Higgs 
in the collision of QCD partons $a$ and $b$ is
%=========================
\begin{equation}
\label{eq:LPfactorization}
d\hat{\sigma}_{ab\rightarrow H+X}(p_a,p_b,P)
= \sum_i \int_0^1 \!d z\, 
d\tilde{\sigma}_{ab\rightarrow i+X}
(p_a,p_b,p_i=\tilde{P}/z;\mu) \,
D_{i\rightarrow H}(z,\mu),
\end{equation}
%=========================
where $p_i$ is the momentum of the fragmenting parton 
and $\tilde{P}$ is a light-like 4-vector whose 3-vector component is $\mathbf{P}$.
The relation $p_i=\tilde{P}/z$ is valid only when the fragmenting parton mass $M_i$ can be ignored.
The effect of nonzero $M_i$ can be systematically considered with a more complicated factorization prescription \cite{Braaten:2015ppa}.
The errors in the factorization formula in Eq.~(\ref{eq:LPfactorization}) are of order $M_H^2/P_T^2$.
The sum over partons $i$ is over the types of elementary particles in the Standard Model. 
The most important fragmenting partons for Higgs production  
are the weak vector bosons $W$ and $Z$, the top quark $t$,  the top antiquark $\bar t$, 
the gluon $g$, and the Higgs itself.
The integral in Eq.~(\ref{eq:LPfactorization})
is over the longitudinal momentum fraction $z$ of the Higgs 
relative to the fragmenting parton.  The integration range of $z$ is from 0 to 1,
up to kinematic constraints from $d\tilde{\sigma}_{ab\to i+X}$.
The fragmentation functions $D_{i\rightarrow H}(z,\mu)$ are distributions for $z$
 that depend on $M_H$
and on the mass  $M_i$ of the fragmenting parton.
The only dependence on $P_T$ in Eq.~(\ref{eq:LPfactorization})
is in the cross sections $d\tilde{\sigma}$ for producing the parton $i$.
In these cross sections, the Higgs mass $M_H$ is set to $0$, while the mass $M_i$ of the fragmenting parton may or may not be at its physical value.
Note that the hard-scattering cross section $d\hat{\sigma}$
on the left side of Eq.~(\ref{eq:LPfactorization})
has mass singularities in the limits $M_H\to 0$ and $M_i \to 0$.
Since the parton production cross sections $d\tilde{\sigma}$ have no mass singularities, 
we will refer to them
as {\it infrared-safe cross sections}.
The infrared-safe cross sections $d\tilde{\sigma}$ and the fragmentation functions 
can all be calculated perturbatively as expansions in powers of $\alpha_s$
and the other coupling constants of the Standard Model.
The separation of the hardest scale $P_T$ from the softer scales $M_H$ and $M_i$
involves the introduction of another arbitrary
factorization scale $\mu$. 

There are important differences between the factorization formula in 
Eq.~(\ref{eq:LPfactorization}) for inclusive Higgs production at large $P_T$
and the analogous QCD factorization formula  for inclusive hadron production. 
One difference is that a Higgs can be produced directly in the hard scattering.
Thus the Higgs is included in the sum over fragmenting partons in Eq.~(\ref{eq:LPfactorization}).
In contrast, a hadron at large $P_T$ cannot be produced directly in the hard scattering.
Another difference is that the fragmentation functions for Higgs production are completely perturbative.
They can be calculated order-by-order in the Standard Model coupling constants. 
In contrast, the fragmentation functions for hadron production are nonperturbative,
although their evolution with the  fragmentation scale is perturbative.

%%%%%%%%%%%%%%%%%%%%%%%%

\section{Fragmentation functions}

One of the most important partons that fragments into the Higgs is the Higgs boson itself.
The LO fragmentation function for Higgs 
into Higgs is a delta function:
%===================
\begin{equation}
\label{eq:DHH}
D_{H\to H}(z)=\delta(1-z) + {\cal O}(g_W^2, y_t^2).
\end{equation}
%===================
The leading corrections  
are of order $g_W^2$ from the coupling of the Higgs to weak vector bosons 
and of order $y_t^2$ from the Yukawa coupling of the Higgs to the top quark. 

The leading-order contribution to the fragmentation function for top quark  
into Higgs  comes from the tree-level process $t^* \to H+t$. 
The fragmentation function
can be expressed as an integral over $Q^2$, the square of the  invariant mass 
of the final state $H+t$, from $M_H^2/z + M_t^2/(1-z)$ to $\infty$.
The integral over $Q^2$ is divergent.
In the $\overline{\mathrm{MS}}$ renormalization scheme, the LO fragmentation function for $t \to H$ is
%========================
\begin{equation}\label{eq:t->H}
D_{t\to H}(z,\mu) =
\frac{y_t^2}{16 \pi^2} \, z
\left( \log\frac{\mu^2}{M_t^2}
- \log\big[z^2+4\,\zeta_t(1-z)\big]
+4\,(1-\zeta_t)\frac{1-z}{z^2+4\,\zeta_{t}\, (1-z)}
\right),
\end{equation}
%=========================
where $\zeta_t \equiv M_H^2/(4M_t^2) \approx 0.13$.
The fragmentation function for $\bar{t} \to H$  is the same.

In Ref.~\cite{Braaten:2015ppa}, the top quark fragmentation function is also calculated in the invariant-mass-cutoff (IMC) scheme, in which the integral over $Q^2$ extends only up to $\mu^2$. The fragmentation functions for the $W$ and $Z$ bosons into Higgs are calculated at LO in both the $\overline{\mathrm{MS}}$ and IMC renormalization schemes. The effect of evolution on all these fragmentation functions is also studied.

%%%%%%%%%%%%%%%%%%%%%%%

\section{Comparison with a complete LO calculation}

In this section, we apply the LP factorization formula
to the process $q\bar{q}\to H t\bar{t}$ at LO, where $q$ is a massless quark. 
At LO, the LP factorization formula in Eq.~(\ref{eq:LPfactorization}) 
for the process $q\bar{q}\to H t\bar{t}$ has two terms:
%=========================
\begin{equation}
\label{eq:fac_formula_LO_PT}
\frac{d^2\hat{\sigma}_{q\bar{q}\rightarrow H t\bar t }}{d P_T^2 d \hat y}
=
\frac{d^2\tilde{\sigma}_{q\bar{q}\rightarrow H+t\bar t}}{d P_T ^2 d \hat y}
+ 2\int_0^1 \frac{d z}{z^2}\, 
\frac{d^2\tilde{\sigma}_{q\bar{q}\rightarrow t+\bar{t}}}{d p_T^2  d \hat y}
(p=\tilde{P}/z) \,
D_{t\rightarrow H}(z).
\end{equation}
%=========================
where $P$ is the momentum of the Higgs with mass $M_H$, 
$\tilde{P}$ is the corresponding momentum for a massless Higgs, and
$p$ is the momentum of the fragmenting top quark.
The subscripts $T$ represent the transverse components, and
$\hat y$ is the Higgs rapidity in the partonic center-of-momentum frame.
In Eq.~(\ref{eq:fac_formula_LO_PT}), we have explicitly used Eq.~(\ref{eq:DHH}) and the fact that the fragmentation contributions 
from $t$ and $\bar{t}$ are the same. The relevant fragmentation function is given in Eq.~(\ref{eq:t->H}).

The infrared-safe cross sections $d^2 \tilde\sigma$ can be calculated using different prescriptions. In Ref.~\cite{Braaten:2015ppa}, we use both the zero-mass-top-quark (ZMTQ) prescription and the hybrid prescription. In the ZMTQ prescription, both the top quark mass and the Higgs mass are set to zero in the infrared-safe cross sections in Eq.~(\ref{eq:fac_formula_LO_PT}). Consequently the error of the LP factorization formula is $\mathcal{O}(M_t^2/P_T^2)$.
In the hybrid prescription, the top quark mass is kept at its physical value in the infrared-safe cross section for $q\bar{q} \to H +t\bar t$. This decreases the error to $\mathcal{O}(M_H^2/P_T^2)$.

%%%%%%%%%%%%%%%%%%%
\begin{figure}[htb]
\centering
\includegraphics[height=1.8in]{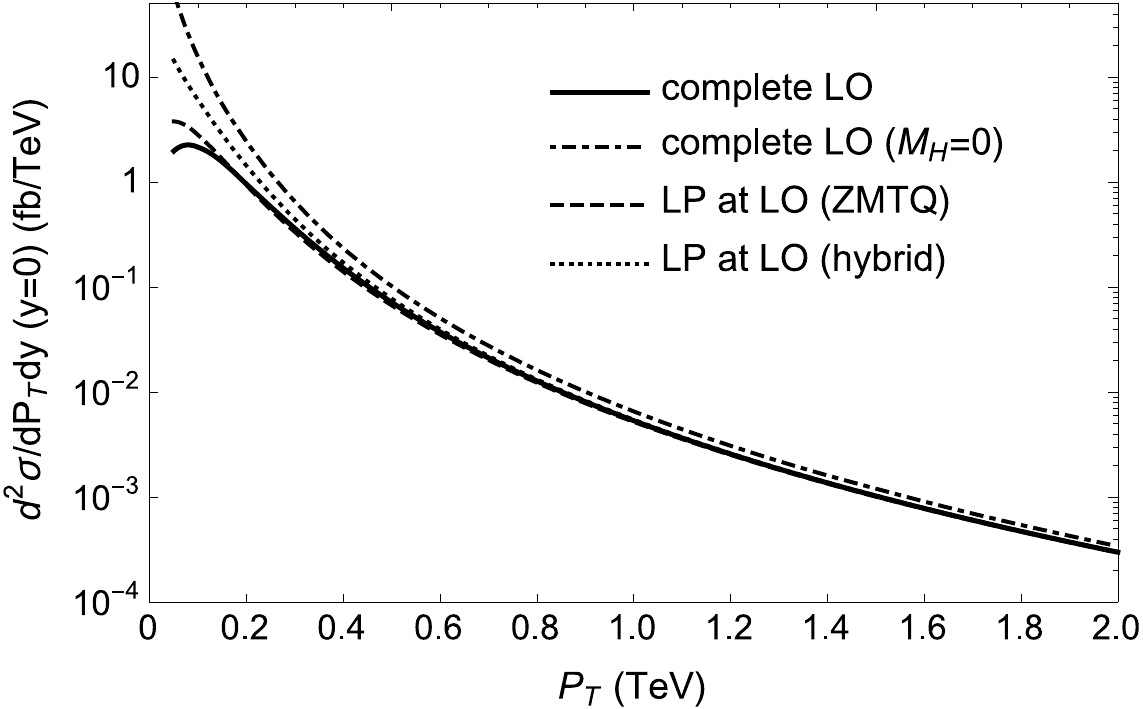}
\includegraphics[height=1.8in]{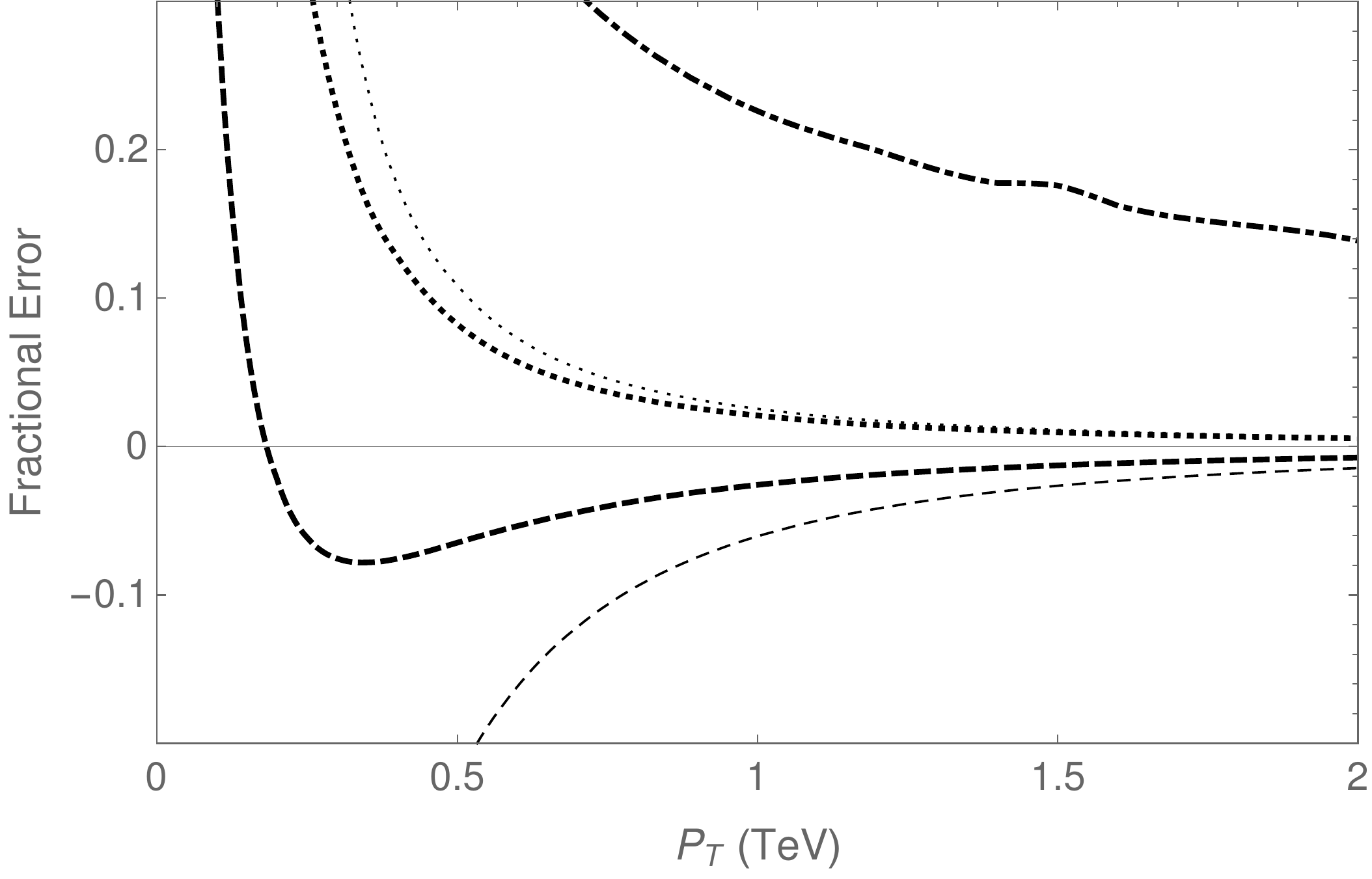}
\caption{Left panel: Differential cross section for inclusive Higgs production at central rapidity
from the parton process $q\bar{q}\to t\bar{t}H$ at a 100~TeV $p p$ collider.
Right panel: Fractional error in the differential cross section for inclusive Higgs production. 
The thick (thin) lines are the LP factorization results in the $\overline{\mathrm{MS}}$ (IMC) renormalization schemes.
See text for more details.}
\label{fig:compare}
\end{figure}
%%%%%%%%%%%%%%%%%%%

On the left panel in Fig.~\ref{fig:compare}, we show the differential cross section for inclusive Higgs production at central rapidity from the parton process $q\bar{q}\to t\bar{t}H$ at a 100~TeV $p p$ collider
as a function of the Higgs transverse momentum $P_T$. The LP factorization results are calculated with top quark fragmentation functions in the $\overline{\mathrm{MS}}$ renormalization scheme. For comparison, we also show another approximation obtained by setting the Higgs mass to zero in the complete LO result.
On the right panel in Fig.~\ref{fig:compare}, we show the fractional error, which is the difference between the approximate result and the complete LO result divided by the complete LO result. The fractional error of the LP factorization formula with the top quark fragmentation function in the IMC renormalization scheme is also plotted.

Fig.~\ref{fig:compare} shows that 
the LP factorization formula with the ZMTQ and hybrid prescriptions
 both give increasingly good approximations  to the complete LO result at large $P_T$,
with the errors decreasing to below $5\%$ for $P_T > 600$ GeV.
 In contrast, the fractional error for the complete LO result 
with $M_H=0$ does not go to zero at large $P_T$.
This result indicates that the LP factorization formula is a systematic method to separate the mass scales from $P_T$, with the error decreasing with $P_T$, while the complete LO with $M_H=0$ is not a good approximation.
More detailed analysis shows that setting $M_H=0$ in the region where Higgs is produced collinearly to the top quark results in an error $\mathcal{O}(M_H^2/M_t^2)$, which does not decrease with increasing $P_T$ \cite{Braaten:2015ppa}.

%%%%%%%%%%%%%%%%%%

\section{Summary and outlook}

We have presented the leading-power (LP) factorization formula for Higgs production 
with transverse momentum $P_T$  much larger than the mass $M_H$ of the Higgs.
We have calculated the LO fragmentation functions for Higgs, top quark, $W$ and $Z$ into Higgs.
By comparing the LP factorization formula with the complete calculation at LO for the process $q\bar{q}\to H t\bar{t}$, we find that the error in the LP factorization formula decreases to less than $5\%$ for $P_T>600$~GeV at a future 100~TeV collider.

The LP factorization formula can be applied to other Higgs production processes, as well as the production of other particles at $P_T$ much larger than their masses.
The separation of scales in the LP factorization formula simplifies the higher order calculations. Moreover, the factorization formula can be extended to
the next-to-leading-power (NLP) \cite{Kang:2011mg, Fleming:2012wy, Kang:2014tta}, reducing the error to $\mathcal{O}(M_H^4/P_T^4)$, which is about 6\% at $P_T=250$~GeV. Thus the NLP factorization formula could be useful even at the Large Hadron Collider.

\end{document}